\documentclass[12pt]{iopart}

\usepackage{graphicx,xcolor}
\expandafter\let\csname equation*\endcsname\relax
\expandafter\let\csname endequation*\endcsname\relax
\usepackage{amsmath, amsthm, amssymb}
\usepackage{float}
\usepackage{subfig}
\usepackage{xcolor}
\usepackage{multirow}
\usepackage{svg}
\usepackage{hyperref}

\UseRawInputEncoding

\begin{document}

\title{Experiment-free disruption prediction for new devices enabled by synthetic diagnostic data augmentation}

\markboth{Experiment-free disruption prediction by synthetic diagnostic data}{}

\author{
Zhiqiang Liu\textsuperscript{1,2},
Fengming Xue\textsuperscript{2},
Shiwei Xue\textsuperscript{2},
Bihao Guo\textsuperscript{4},
Dalong Chen\textsuperscript{4},
Wei Zheng\textsuperscript{2,*},
Ping Zhu\textsuperscript{2,3,*}
and J-TEXT team\textsuperscript{2}
}

\address{
\textsuperscript{1}
School of Physics, Huazhong University of Science and Technology,
Wuhan, Hubei 430074, China
}

\address{
\textsuperscript{2}
State Key Laboratory of Advanced Electromagnetic Technology, \\
International Joint Research Laboratory of Magnetic Confinement Fusion
and Plasma Physics, School of Electrical and Electronic Engineering, \\
Huazhong University of Science and Technology,
Wuhan, Hubei 430074, China
}

\address{
\textsuperscript{3}
Department of Nuclear Engineering and Engineering Physics, \\
University of Wisconsin-Madison,
Madison, WI 53706, United States of America
}

\address{
\textsuperscript{4}
Institute of Plasma Physics, CAS, PO Box 1126, Hefei 230031, China
}

\ead{zhup@hust.edu.cn, zhengwei@hust.edu.cn}

\vspace{10pt}
\begin{indented}
\item[]May 14th, 2026
\end{indented}
\clearpage

\begin{abstract}

Deep learning based approaches have shown great promise in cross-device disruption prediction for tokamaks, however, the robustness of these models heavily relies on massive amounts of training data. For the upcoming ITER, to ensure the safety of the first plasma and subsequent operations, experimental data should be entirely unavailable initially, and disruptive discharges should be strictly avoided thereafter. This extreme data scarcity inherently conflicts with the data-intensive nature of deep learning algorithms. To address this challenge, we utilize synthetic diagnostic signals from the target device to supplement the experimental data from existing devices for the zero-shot disruption prediction on a new device. The detailed implementation pipeline of this scheme is presented. For experimental validation, a predictive model trained on data from the EAST tokamak is deployed for a zero-shot cross-device experiment on the J-TEXT tokamak. A synthetic diagnostic framework, configured with the diagnostic parameters of the target device, is developed to process NIMROD magnetohydrodynamic (MHD) simulation data based on the target device's magnetic configuration, thereby achieving effective data augmentation. Ultimately, the results demonstrate that by integrating the target device's synthetic diagnostic data with Fourier Domain Adaptation, the zero-shot accurate early warning rate of the model on 1,596 J-TEXT discharges is improved from 50\% to 57\%, while exhibiting enhanced predictive robustness. 

\end{abstract}

\vspace{2pc}
\noindent{\it Keywords}: Disruption prediction, Synthetic diagnostics, Domain adaption

\submitto{\PPCF}
\maketitle
\section{Introduction}

Tokamaks currently stand as the most promising magnetic confinement configuration for achieving controlled neuclear fusion . However, the operation of tokamaks is often threatened by major disruptions. These events represent rapid losses of plasma confinement that lead to extreme thermal and electromagnetic loads on device structures\cite{JET_SurveyDisruptionCauses2011}. As the international fusion community approaches the first plasma of the ITER project, ensuring the safety of such a massive device has become an urgent challenge. Unlike current experimental devices where disruptions are often tolerated as a byproduct of exploring operational boundaries, ITER cannot accommodate disruptive operations that risk structural integrity. According to its stringent operational limits, the disruption rate must be kept strictly below 1\% during full-power operations\cite{ITER_DisruptionPrevention2019}. Therefore, developing a highly accurate, robust, and reliable early warning system for disruptions is not merely an optimization problem, but an absolute prerequisite for the success of DEMO\cite{parkMainNuclearResponses2024}, as well as future commercial fusion reactors.\\

In recent years, deep learning (DL) has driven a paradigm shift in real-time disruption prediction. Neural networks have demonstrated exceptional performance in capturing non-linear relationships between multi-dimensional diagnostic signals and impending physical instabilities \cite{HL2A_Diruption_prediction2023,zheng_nc_2023,seoAvoidingFusionPlasma2024}. Yet, despite their success on individual devices, these models consistently suffer from severe performance degradation when transferred from the training source device to a new target device. This Out-of-Distribution (OOD) \cite{OOD_OutofdomainGeneralization2022c} generalization problem stems from two main sources. First, at the data distribution level, discrepancies in experimental diagnostics, including differences in sensor hardware, spatial configurations, and noise profiles, alter the marginal distribution of the input signals. Second, at the physics level, the mechanisms triggering disruptions in next-generation devices like ITER are far more complex. These include novel disruption causes such as alpha-particle heating effects and distinct impurity accumulation behaviors, which lack latent feature representations in historical datasets \cite{montesMachineLearningDisruption2019}. Such intrinsic discrepancies fundamentally limit the direct transferability of DL models.\\

To mitigate these OOD issues, current cross-device transfer learning paradigms have evolved through several innovative approaches: one strategy establishes cross-regime mapping by leveraging the correlation between low-performance and high-performance discharges to bridge the operational gap within target devices \cite{zhu_ScenarioAdaptive2021}. Another approach utilizes physics-guided feature extraction and improved domain adaptation algorithm to reduce the demand for target tokamak discharges \cite{shen_PGFE2023}. Alternatively, adaptive anomaly detection frameworks based on enhanced convolutional autoencoders (E-CAAD) have been proposed for cross-tokamak disruption prediction, enabling prediction from the first discharge on a new device through adaptive transfer and continual retraining strategies  \cite{ai_AdaptiveAnomalyDetection2025}. And recent efforts focus on extracting device-independent temporal representations from massive multi-machine datasets, demonstrating that while stable states are machine-specific, the dynamical precursors to disruptions follow universal physical patterns \cite{zhu_hybridTokamak2021}.
While these schemes show promise, they do not resolve the absolute data absence problem. They introduce an irreconcilable paradox for next-generation large-scale projects: DL models inherently rely on vast amounts of annotated disruption samples to ensure robustness, whereas ITER will completely lack experimental disruption data during its initial stages and must strictly avoid disruptive discharges subsequently. Thus, the model's efficacy under future ITER operations remains unguaranteed. Achieving accurate zero-shot cross-device prediction, which involves deploying a model without requiring any target device disruption data, has become a frontier in fusion research.\\

Although  the simulation-assisted framework have already been attempted for tokamak plasma control and state reconstruction, their application to zero-shot disruption prediction remains largely unexplored. Previous studies employed physics-based simulators such as RAPTOR together with synthetic diagnostic signals to support equilibrium reconstruction and real-time feedback control in devices like TCV \cite{ControlSimulations2017}. The work demonstrated that synthetic diagnostics \cite{shiSyntheticDiagnosticsPlatform2016} can effectively bridge plasma physics models and experimental systems. However, existing approaches primarily focus on plasma profile regulation and control-oriented state estimation, rather than disruption precursor learning and cross-device disruption generalization. Motivated by these developments, we extend the simulation assisted paradigm toward disruption prediction by introducing synthetic diagnostic data augmentation and sim-to-real \cite{wangEmbodieDreamerAdvancingReal2Sim2Real2025} domain alignment for zero-shot transfer.

\begin{figure}
    \centering
    \includegraphics[width=0.8\textwidth]{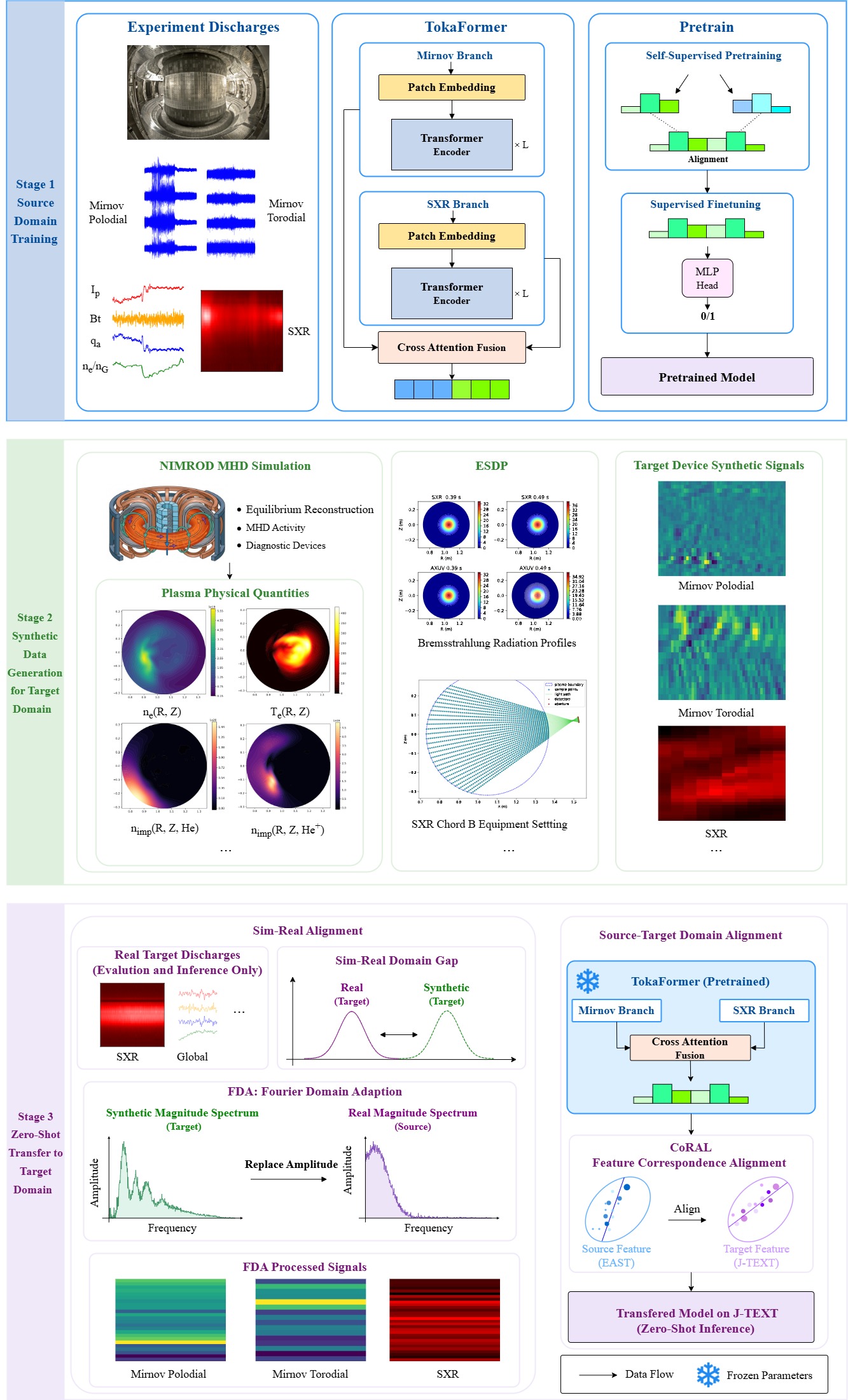}
    \clearpage
    \caption{Overview of the proposed zero-shot cross-device transfer framework for tokamak disruption prediction. }
    \label{fig:Pipeline}
\end{figure}

To break this data scarcity obstacle, we develop a new scheme based on synthetic diagnostic data augmentation to improve the model's zero-shot performance on new devices. The overall architecture of the proposed framework is illustrated in Fig.~\ref{fig:Pipeline}, which consists of source-domain pretraining, synthetic target-domain signal generation, and zero-shot transfer with domain alignment. Advanced 3D non-ideal magnetohydrodynamic (MHD) \cite{Chapter3MHD2007} codes, such as M3D-C1 \cite{m3dc1_2022}, JOREK \cite{JOREK_hoelzlNonlinearMHDModelling2024} and NIMROD \cite{nimrod_spi_2022}, can simulate various disruption events including tearing modes and locked modes under the precise magnetic configurations of target devices \cite{jiang_SMBI2020}. However, simulated data and experimental data often adopt different physical quantities and numerical formats. Simulated data typically consist of spatio-temporal continuous physical fields such as electron density and temperature, whereas experimental data comprise discrete temporal signals constrained by hardware such as interferometers or Mirnov coils. Based on diagnostic physics, we built Enhanced Synthetic Diagnostic Platform (ESDP) which can bridge this gap by computationally projecting the simulated physical fields through the specific diagnostic layout and parameters of the target device, thereby generating synthetic experimental signals.  To rigorously validate the feasibility of this zero-shot transfer scheme, we utilize the EAST \cite{EAST_OverviewRecentExperimental2024} superconducting tokamak as the source device and the J-TEXT \cite{dingOverviewRecentExperimental2024} conventional tokamak as the target device. The significant differences in their geometric dimensions, operational regimes, and diagnostic layouts provide an ideal testbed for evaluating cross-device generalization.\\

According to the domain adaptation theory \cite{ben-david_TheoryLearningDifferent2010}, the upper bound of the target domain error is strictly constrained by both the source domain error and the cross-domain feature distance. To minimize the source error, we introduce a dual-tower network named TokaFormer, which is engineered to independently process high-frequency Mirnov signals and low-frequency macroscopic parameters including soft X-ray emission (SXR) \cite{J-TEXT_SoftXrayImaging2009}, plasma current $I_p$, toroidal magnetic field $B_t$, Greenwald density fraction ($\bar{n_e}$/$n_G$ ), and edge safety factor ($q_a$). Furthermore, to provide better prior knowledge for the fine-tuning of the disruption prediction stage, we employ a self-supervised Image-Text Contrastive (ITC) pre-training scheme. By learning from an extensive ensemble of experimental discharges, the model develops highly robust feature representations of plasma states, effectively enhancing its transferable predictive capabilities. 
Regarding the latter constraint of mitigating the cross-domain distance, the absolute scarcity of target experimental data poses a critical challenge, given that idealized synthetic diagnostics inherently introduce a severe simulation-to-reality (sim-to-real) domain shift. To safeguard the cross-device transfer from being confounded by this discrepancy, we propose a hierarchical two-stage alignment protocol. Specifically, Fourier Domain Adaptation (FDA) \cite{FDA_2020} is first adopted to substitute the amplitude spectrum of the target synthetic data with that of source experimental discharges, while strictly preserving the phase spectrum that encapsulates the underlying MHD dynamics.
With the sim-to-real shift mitigated by FDA, deep Correlation Alignment (CORAL) \cite{sunDeepCORALCorrelation2016} is subsequently executed to match the second-order covariance structures in the latent feature space. This novel joint mechanism effectively minimizes the cross-device feature distance, closing the cross-domain gap between the source and target experimental distributions.\\

The rest of the article is organized as follows. Section II introduces the preparation of the experimental datasets from the EAST and J-TEXT  tokamaks. It also describes the MHD simulations conducted using the NIMROD code and presents the ESDP developed to generate synthetic diagnostic signals. Section III gives details on the zero-shot cross-device transfer methodology. This section presents the multi-modal dual-tower architecture of TokaFormer, elucidates the self-supervised pre-training paradigm governed by the ITC loss, and demonstrates the hierarchical implementation of FDA and CORAL designed to sequentially resolve the simulation-to-reality domain shift and the cross-device feature divergence. Finally, Section IV summarizes the study and provides the conclusion.

\section{Experiment Dataset and Synthetic Signals}
\label{dataset}

\subsection{Dataset Preprocessing}
EAST is a fully superconducting tokamak operating primarily in a divertor configuration, with a major radius $R = 1.85~\mathrm{m}$, a minor radius $a = 0.45~\mathrm{m}$, a plasma current $I_p \sim 450~\mathrm{kA}$, and a toroidal magnetic field $B_t \sim 1.5~\mathrm{T}$. Typical discharges on EAST are characterized by a normalized beta $\beta_N \sim 2.1$ and extended pulse lengths of approximately 10~s. In contrast, J-TEXT is a medium-sized tokamak operating in a limiter configuration with a more compact geometry, characterized by major radius $R = 1.05~\mathrm{m}$ and minor radius $a = 0.25~\mathrm{m}$. Its typical operational space includes $I_p \sim 200~\mathrm{kA}$, $B_t \sim 2.0~\mathrm{T}$, and shorter pulse lengths ranging from 700 to 800~ms~\cite{shen_coral}.

To ensure cross-device consistency, we selected diagnostic signals common to both J-TEXT and EAST as model inputs, which include those from SXR arrays and poloidal and toroidal Mirnov arrays, as well as $I_p$ and $B_t$. Additionally, derived physical parameters are included to provide macroscopic constraints, namely the Greenwald density fraction
\begin{equation}
\frac{\bar{n_e}}{n_G} = \frac{\bar{n_e} \pi a^2}{I_p},
\end{equation}
and the edge safety factor
\begin{equation}
q_a = \frac{2\pi a^2 B_t}{\mu_0 R I_p}.
\end{equation}

To address the issue of missing information caused by faulty diagnostic channels, a second-order interpolation method is employed to reconstruct the corrupted signals \cite{data_clean_2026}. The Z-score normalization,
\begin{equation}
x' = \frac{x - \bar{x}}{\sigma},
\end{equation}
is applied to the mitigation of magnitude discrepancies among heterogeneous diagnostic signals. For the source device, the global mean and variance of each diagnostic are calculated across the entire dataset. For the target device, the normalization parameters are derived from the synthetic diagnostic signals.

The labeling strategy and data sampling are meticulously designed to synchronize the temporal features with the physical scales of each device. For disruptive samples, we extract a continuous observation window immediately preceding the current quench (CQ)~\cite{zeng_current_spike2023} to capture the disruption precursors. The duration of this window is determined by the characteristic resistive time scale
\begin{equation}
\tau_R = \frac{a^2}{\eta},
\end{equation}
which scales with the device’s minor radius $a$ and the plasma resistivity ${\eta}$. For EAST, where the typical $\tau_R$ exceeds 500~ms, we define a 125~ms precursor window. In contrast, for J-TEXT, $\tau_R$ is significantly shorter at approximately 25~ms, thus a 25~ms window is adopted.

To prevent the model from suffering from severe class imbalance, a common issue in tokamak datasets where non-disruptive phases dominate, we implement a balanced downsampling strategy for negative samples using the $I_p$ down time as the dividing boundary. Specifically, the non-disruptive dataset is constructed from two sources: (i) continuous time windows of the same duration taken prior to the termination of non-disruptive discharges, and (ii) discrete segments randomly sampled from the flattop phases of both disruptive and non-disruptive pulses. We maintain a 3:1 ratio between negative and positive samples to balance operational realism with training stability, utilizing a dynamic resampling strategy during training to ensure sufficient sensitivity to rare disruptive events.

\subsection{NIMROD MHD Simulation}

To generate the necessary training data for the target device, we perform MHD simulations using the NIMROD code, based on the single-fluid model configured with J-TEXT parameters. Specifically, the simulation setup for impurity-induced instabilities follows the parameters established in the previous study of Massive Gas Injection (MGI)~\cite{zeng_cold_bubble2022}. Additionally, to simulate stable flattop phases and generate negative samples, an initial magnetic perturbation of approximately 5~Gauss is introduced to produce quasi-steady-state signals without triggering a disruption.

Given the high computational cost of first-principle MHD simulations, we have implemented several strategies to maximize data utilization and perform physics-informed data augmentation:
\begin{enumerate}
    \item \textbf{Optimization of impurity injection level:} Excessive impurity level can lead to an unnaturally rapid pre-thermal quench (pre-TQ) phase, limiting the availability of useful precursor data. We therefore have adjusted the impurity density to $3.1 \times 10^{18}~\mathrm{m}^{-3}$, a level used in Supersonic Molecular Beam Injection (SMBI) simulation to trigger minor disruptions that eventually evolve into major ones~\cite{jiang_SMBI2020}, thereby capturing richer and more varied disruption evolution stages.
    \item \textbf{Domain randomization via equilibrium profiles:} We utilize diverse discharge equilibrium configurations as initial plasma conditions to simulate diagnostic signal evolution across different operational regimes, effectively enhancing the model's generalization ability across various plasma states.
    \item \textbf{Toroidal angle augmentation:} Although experimental diagnostic systems are typically fixed at certain toroidal positions, we can strategically exploit the simulated toroidal variations to implement a physically consistent data augmentation technique. Specifically, plasma profiles are extracted from a single 3D NIMROD simulation case at $30^\circ$ toroidal angle intervals to compute subsequent synthetic diagnostic signatures. This procedure yields 12 distinct training samples with subtle physical variations from each simulation run, substantially increasing data diversity while strictly adhering to the underlying plasma physics.
\end{enumerate}

\subsection{Synthetic Diagnostics}

To bridge the gap between numerical simulations and experimental observations, we developed the ESDP\cite{esdp2025software} framework based on the SDP codebase\cite{shiSyntheticDiagnosticsPlatform2016} by developing and integrating four specialized diagnostic modules: the polarimeter-interferometer, soft X-ray (SXR), absolute extreme ultraviolet (AXUV), and Mirnov coils. This framework takes plasma profiles, including electron density ($n_e$), ion density ($n_i$), impurity density ($n_{\mathrm{imp}}$), magnetic field ($B$), electron temperature ($T_e$), and ion temperature ($T_i$) from numerical simulations as input. By integrating diagnostic physics, ESDP projects these physical quantities onto the diagnostic layout of a target device to generate synthetic experimental signals.

In this study, the geometric and instrumental parameters, such as laser wavelength, responsivity, and circuit gains, within each ESDP module are adjusted to match the instruments on J-TEXT. We then process NIMROD simulation data to generate the synthetic SXR, Mirnov, and polarimeter-interferometer system (POLARIS) \cite{polaris2022} signals required for cross-device transfer. The modeling procedures are as follows:

\textbf{i) SXR arrays:} The radiation emissivity profiles, encompassing bremsstrahlung, recombination, and line radiation, are first calculated from the simulated plasma profiles via the following equations \cite{radiation_equations2023a}. The bremsstrahlung power density is calculated as:
\begin{equation}
P_{\mathrm{brem}}
=
2.09 \times 10^{-36}
\, g Z^2
\left(
\frac{n_e n_i}{\lambda_s^2 T_e^{1/2}}
\right)
\exp
\left(
-\frac{1.24 \times 10^4}{\lambda_s T_e}
\right)
\label{eq:brem}
\end{equation}
where $n_i$ denotes the ion density, $g$ is the Gaunt factor, $Z$ is the ion charge number, and $\lambda_s$ represents the radiation wavelength. 
For recombination radiation, the power density is given by:
\begin{equation}
P_{\text{rec}} = 1.6 \times 10^{-19} R_{\text{rec},c} n_c \left( E_{\text{ion},c} + T_e \right)
\label{eq:rec_power}
\end{equation}
where \(E_{\text{ion},c}\) is the energy level difference, and \(R_{\text{rec}}\) is the recombination rate coefficient of electrons and ions, which can be calculated by the following formula:
\begin{align}
R_{\text{rec},c} &= 5.2 \times 10^{-14} (c+1) n_e f_{\text{rec}} \sqrt{\frac{E_{\text{ion},c}}{T_e}}
\label{eq:rec_rate} \\
f_{\text{rec}} &= 0.43 + \frac{1}{2} \log_{10}\left( \frac{E_{\text{ion},c}}{T_e} \right) + 0.469 \sqrt[3]{\frac{E_{\text{ion},c}}{T_e}}
\label{eq:f_rec}
\end{align}
where \(c\) denotes the impurity charge state, ranging from 0 to the nuclear charge number \(Z\) of the impurity.
Similarly, the line radiation power density is defined as:
\begin{equation}
P_{line} = 10^{\alpha(T_e)} \times 10^{-13} n_e n_{Z,(c-1)}
\label{eq:line_power}
\end{equation}

\begin{equation}
\alpha(T_e) = \sum_{c} \alpha_c(T_e) \left( \log_{10} \frac{T_e}{1000} \right)^{c-1}
\label{eq:alpha_Te}
\end{equation}
where $n_{Z,(c-1)}$ is the electron density of the impurity at charge state $(c-1)$, and $\alpha(T_e)$ is the line radiation coefficient as a function of $T_e$, which can be obtained from ADAS (Atomic Data and Analysis Structure) \cite{openadas2025}.\\

In practical radiation calculations, the spectral response of the detector must also be considered. On J-TEXT, the front-end of the SXR detector is equipped with a beryllium (Be) window of specific thickness, which filters out low-energy ultraviolet and visible radiation below 500\,eV, ensuring that the detector primarily responds to high-energy X-ray photons emitted from the hot core plasma region.
Then the signal intensity for each channel is finally determined by
\begin{equation}
I = R_{\mathrm{res}} \eta
\frac{A_{\mathrm{ph}} A_{\mathrm{ap}} \cos\theta_{\mathrm{ph}} \cos\theta_{\mathrm{ap}}}{4\pi d^2}
\int \epsilon(R,Z)\, dL.
\label{eq:sxr}
\end{equation}
where \(R_{\text{res}}\) is the responsivity of the photodiode, \(A_{\text{ph}}\) and \(A_{\text{ap}}\) are the physical areas of the photosensitive element and the aperture, respectively, \(\cos\theta_{\text{ph}}\) is the cosine of the angle between the detector and the viewing chord, and \(d\) is the distance between the photosensitive element and the aperture.

\textbf{ii) Mirnov arrays:} Based on Faraday’s law, synthetic voltage signals are derived by calculating the time derivative of the magnetic flux through the effective area of the pick-up coils, oriented according to the actual J-TEXT sensor distribution.

\textbf{iii) POLARIS:} The synthetic POLARIS diagnostic is modeled to simultaneously compute the Faraday rotation angle and the line-integrated electron density. These physical quantities are evaluated by integrating the simulated electron density and the magnetic field along the specific laser beam paths, accounting for the emission wavelength of the target device.

\begin{equation}
\alpha = \frac{\Delta\phi_R - \Delta\phi_L}{2} = c_p \int n_e(z) B_{\parallel}(z) dz
\label{eq:3-1}
\end{equation}

\begin{equation}
\varphi_{n_e L} = \frac{\Delta\phi_R + \Delta\phi_L}{2} = c_i \int n_e(z) dz
\label{eq:3-2}
\end{equation}
where
\[
c_p = \frac{e}{2 m_e c n_c} = 2.62 \times 10^{-13} \lambda^2, \quad
c_i = \frac{2\pi}{\lambda n_c} = 2.81 \times 10^{-15} \lambda,
\]
and $\lambda$ is the emission laser wavelength of the diagnostic system.

\begin{figure}
    \centering
    \includegraphics[width=\textwidth, height=0.9\textheight, keepaspectratio]{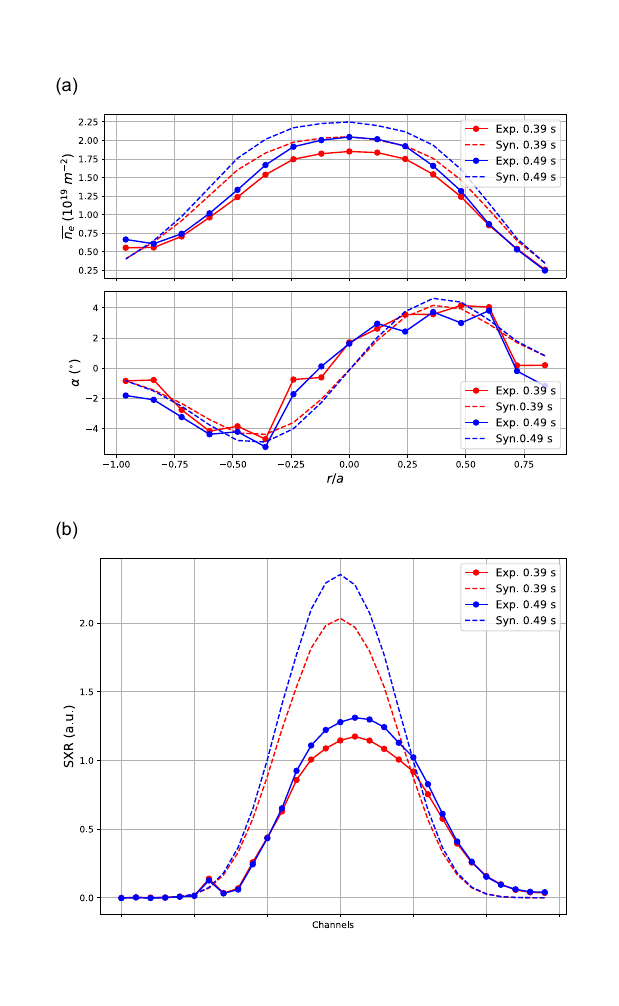}
    \caption{Comparison of POLARIS and SXR signals between experimental measurements (solid lines with markers) and synthetic results (dashed lines) for J-TEXT discharge \#1070019 at 390~ms and 490~ms. The synthetic signals are generated using the EFIT reconstructed equilibrium.}
    \label{fig:sxr_compare}
\end{figure}

To fully validate the effectiveness of the synthetic diagnostic modules, we use the physical profiles reconstructed by EFIT at two different moments during the discharge of J-TEXT shot \#1070019. Based on these profiles, we calculate the synthetic diagnostic results for the same times and compared them with the experimental measurements. 
The detailed comparison results are shown in in Fig.~\ref{fig:sxr_compare}, where the synthetic SXR profiles exhibit an offset compared to the experimental data. This discrepancy primarily arises from the inherent limitations of EFIT reconstruction, which lacks impurity distribution profiles. Consequently, the calculated synthetic emissivity is merely dominated by bremsstrahlung, whereas actual experimental signals include impurity contributions such as line radiation and recombination radiation. Despite this magnitude difference, the spatial distributions and temporal trends remain physically consistent, ensuring that the synthetic signals serve as effective features for model transfer.

To ensure compatibility with disruption prediction, the simulated signals are carefully curated. In experimental datasets, disruption precursor samples are defined by a time window preceding the CQ. For the synthetic disruptive signals derived from MGI simulations, we utilize the abrupt spike in the central SXR channel followed by a rapid collapse as the physical criterion to define the onset of disruption. This spike typically signifies the thermal quench (TQ)~\cite{masiIntegrationOpticalSensors2025}, where the core thermal energy is abruptly lost, invariably leading to a subsequent CQ and plasma disruption in actual discharges. Accordingly, the simulation data is truncated at this point, with the phase preceding the TQ used as disruptive precursor samples for the target device. Conversely, simulations involving only magnetic perturbations that do not trigger a thermal collapse are treated as non-disruptive samples. This rigorous partitioning assists the model in distinguishing between disruptive MHD instabilities and quiescent plasma fluctuations on the target device during the transfer learning process.

\section{Zero-shot Cross-device transfer via synthetic diagnostics}
\label{main}

\subsection{TokaFormer: A Dual-tower Architecture for Multi-modal Fusion}

Tokamak diagnostics present a unique challenge for deep learning due to their heterogeneous temporal resolutions. For instance, Mirnov signals typically require high-frequency sampling (e.g., 20~$\mu$s), whereas macroscopic parameters like $I_p$ are often recorded at lower resolutions (e.g., 1~ms). Traditional approaches often rely on data-level fusion via upsampling, which can introduce artifacts and fail to capture the high-level semantic correlations between different physical processes. To address this, we propose \textit{TokaFormer}, a dual-tower architecture designed for feature-level fusion. This post-fusion strategy ensures that the coupling between macroscopic equilibrium shifts and microscopic magnetic fluctuations is captured at a high level of abstraction.

\begin{figure}
    \centering
    \includegraphics[width=\textwidth]{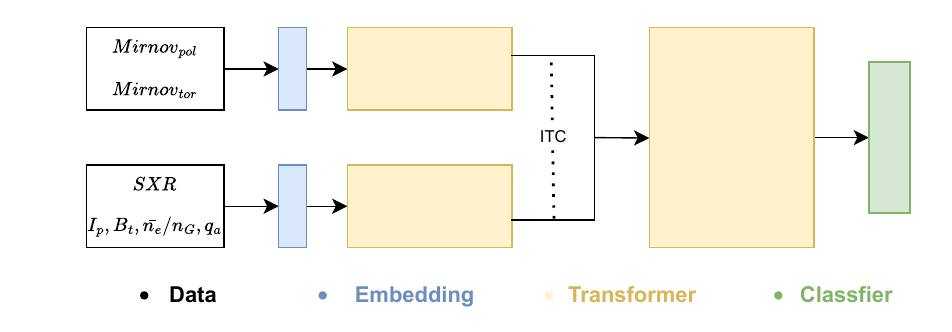}
    \caption{The dual-branch architecture of the TokaFormer model. Inputs are bifurcated into an Electromagnetic Fluctuation Branch (Mirnov signals) and a Thermal Equilibrium Branch (SXR and global parameters: $I_p$, $B_t$, $\bar{n_e}/n_G$, $q_a$). Following independent convolutional embedding and Transformer encoding, the extracted features are fused and fed into a final classification layer for disruption prediction.}
    \label{fig:tokamodel}
\end{figure}

As illustrated in Fig.~\ref{fig:tokamodel}, the model bifurcates the input into two distinct branches based on their physical significance. The \textit{Electromagnetic Fluctuation Branch} is dedicated to Mirnov signals, which are traditionally analyzed via spatial Fourier decomposition of toroidal and poloidal arrays to identify specific $m/n$ perturbation modes. By processing these signals, this branch extracts semantic features associated with magnetic island growth and MHD instability evolution. On the other hand, the \textit{Thermal Equilibrium Branch} integrates $I_p$, $B_t$, $n_G$, $q_a$, with SXR arrays to monitor operational boundaries and plasma states. This design allows the model to identify disruption precursors of current-limit and density-limit while capturing critical information regarding impurity accumulation, radiative cooling, and spatial asymmetries in plasma emissivity.

In terms of architectural realization, 1D convolutional (Conv1D)~\cite{huangDenselyConnectedConvolutional2017} layers are employed as temporal embedding modules to encode the local evolution of SXR and Mirnov array signals within each observation window. Meanwhile, a dense network maps global parameters into embedding vectors. These extracted features are subsequently processed by independent Transformer encoders~\cite{transformer_AttentionAllYou2023} to distill modality-specific physical semantics. To integrate these diverse features, the outputs of both branches are concatenated along the channel dimension. The fusion layer then utilizes a classification token [CLS]~\cite{devlinBERTPretrainingDeep2019} as the unified representation, which is fed into the output classification layer for final disruption prediction.

To evaluate the efficacy of the dual-tower architecture, we conduct comparative experiments against a single-tower baseline under varying model capacities. Both models were trained from scratch using 2,093 EAST discharges and subsequently evaluated on a test set of 200 unseen pulses (comprising 100 disruptive and 100 non-disruptive discharges). To ensure a rigorous assessment, the training hyperparameters, including the learning rate ($1 \times 10^{-4}$), optimizer (Adam), and batch size (128), are kept identical across all trials. The experimental variables were strictly confined to the model's structural parameters, specifically the embedding dimension and the number of Transformer layers.

\begin{table}
\caption{Performance comparison between the baseline (single-tower) and TokaFormer (dual-tower) architectures on the EAST dataset.}
\centering
\begin{tabular}{l c c c c c}
\hline
Experiment & Embedding dimension & Layers & TPR & FPR & Accuracy \\
\hline
Baseline   & 32 & 4 & 0.920 & 0.200 & 0.860 \\
Baseline   & 32 & 6 & 0.920 & 0.180 & 0.870 \\
Baseline   & 64 & 4 & 0.900 & 0.130 & 0.885 \\
TokaFormer & 32 & 4 & 0.920 & 0.150 & 0.885 \\
TokaFormer & 32 & 6 & 0.920 & 0.140 & 0.890 \\
TokaFormer & 64 & 4 & 0.900 & 0.070 & 0.915 \\
\hline
\end{tabular}
\label{tab:architecture}
\end{table}

The predictive performance of the proposed TokaFormer is compared against a single-tower baseline in Table~\ref{tab:architecture}. Evaluated across varying embedding dimensions and Transformer layer depths, the dual-tower architecture demonstrates a clear advantage. Notably, TokaFormer generally achieves same true positive rate (TPR) but lower false positive rates (FPR) and overall accuracy compared to the baseline under identical structural constraints. These findings confirm that the dual-tower architecture provides a more accurate representation of the plasma state, validating the effectiveness of processing electromagnetic and thermal equilibrium modalities through dedicated branches.

\subsection{Unsupervised Pre-training via Inter-modal Alignment}

The scarcity of disruptive events relative to the long duration of stable flattop phases often leads to severe class imbalance in tokamak datasets, with non-disruptive samples exceeding 99\% in the EAST database. Traditional methods often mitigate this by discarding the majority of stable flattop segments to balance the classes. However, this approach omits the foundational physical information embedded in stable discharges, such as the equilibrium constraints and the characteristic spatio-temporal correlations between different diagnostic signals. These signals represent the undisruptive plasma state, which provides a critical reference frame for identifying the onset of instabilities.

To effectively utilize these extensive stable segments, we implement a self-supervised pre-training strategy using ITC, inspired by the Contrastive Language-Image Pre-training (CLIP) framework~\cite{radfordLearningTransferableVisual2021d}. Instead of relying on human-labeled disruption precursors, the model is tasked with aligning the latent representations of the Electromagnetic Fluctuation Branch and the Thermal Equilibrium Branch. In each training batch, the Mirnov features ($q$) and SXR-global features ($k$) are projected into a unified embedding space. The model is optimized to maximize the cosine similarity between paired modalities from the same time segment while minimizing it for mismatched pairs. This dual-tower alignment is governed by the InfoNCE loss function:
\begin{equation}
\mathcal{L}_q = - \log
\frac{
\exp\left(\frac{q \cdot k^{+}}{\tau}\right)
}{
\sum_{i=0}^{N} \exp\left(\frac{q \cdot k_i}{\tau}\right)
},
\label{eq:infonce}
\end{equation}
where $N$ is the batch size, $k^{+}$ is the positive sample from the complementary modality, and $\tau$ is a temperature hyperparameter set to 0.07.

\begin{figure}
    \centering
    \includegraphics[width=\textwidth]{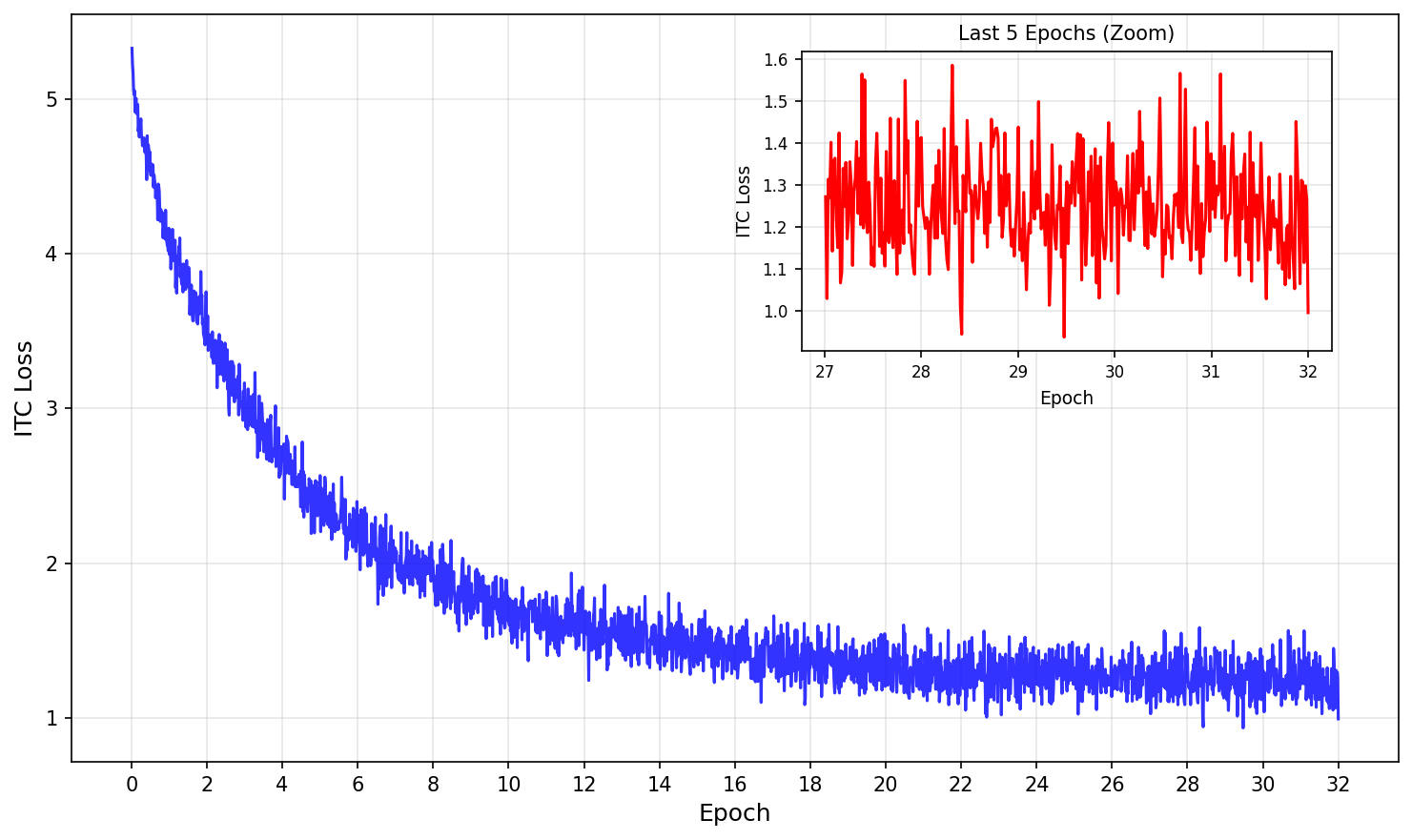}
    \caption{ITC training loss as a function of epoch for the image-text contrastive loss pre-training. The steady convergence of the contrastive loss indicates successful cross-modal alignment between magnetic perturbation and core radiation features across epochs.}
    \label{fig:pretrain_loss}
\end{figure}

By training on a massive dataset of 18,368,382 segments from EAST, the encoders learn to recognize the intrinsic physical coupling between magnetic perturbations and core radiation profiles without explicit supervision. The convergence behavior of the contrastive loss is shown in Fig.~\ref{fig:pretrain_loss}.
The loss exhibits a rapid initial decline as the model captures coarse-grained cross-modal correspondences, followed by a steady stabilization after approximately 20 epochs. This convergence indicates that the model has successfully learned to map disparate diagnostic signals into a consistent feature space.

Following pre-training, we fine-tune the model on the disruption dataset. During this stage, the parameters of the feature encoders are frozen, and only the fusion layers and classification head are updated. This frozen-encoder strategy ensures that the model preserves the high-level physical representations learned during pre-training while adapting the final classification logic to the specific precursors of disruptions.

\begin{table}
\caption{Performance comparison between the model trained from scratch (\textit{Train}) and the pre-trained model (\textit{Finetune}).}
\centering
\begin{tabular}{l c c c c c c}
\hline
\multirow{2}{*}{Experiment} & \multicolumn{3}{c}{EAST} & \multicolumn{3}{c}{J-TEXT} \\
\cline{2-7}
 & TPR & FPR & Accuracy & TPR & FPR & Accuracy \\
\hline
Train & 0.91 & 0.10 & 0.91 & 0.99 & 0.91 & 0.26 \\
Finetune & 0.91 & 0.09 & 0.91 & 0.79 & 0.56 & 0.50 \\
\hline
\end{tabular}
\label{tab:pretrain}
\end{table}

As summarized in Table~\ref{tab:pretrain}, although the model trained from scratch (\textit{Train}) achieves high performance on the source device (EAST), its generalization to the target device (J-TEXT) is poor, evidenced by a significantly high false positive rate (FPR) of 0.91 and a low accuracy of 0.26. This suggests that the scratch-trained model overfits to device-specific noise patterns in the EAST dataset. In contrast, the pre-trained and fine-tuned model (\textit{Finetune}) exhibits a more balanced performance. Although it shows a slight decrease in EAST accuracy, it achieves a substantial improvement in the J-TEXT zero-shot test, raising the accuracy to 0.50 and reducing the FPR to 0.56. These results indicate that the self-supervised alignment of diagnostic modalities helps the encoders distill more universal physical features that are less sensitive to device-specific diagnostic configurations, thereby enhancing the model's robustness for cross-device deployment.

\subsection{Fourier Domain Adaptation and Feature Alignment}

Even with the high-fidelity ESDP modeling, a significant discrepancy persists between the synthetic signals and the experimental observations. Whereas synthetic signals derived from NIMROD simulations represent the idealized physical evolutions, the realistic data from J-TEXT discharges are often embedded in complex background noise and instrumental artifacts. Furthermore, experimental diagnostics are inherently subject to finite bandwidth limitations and measurement uncertainties, which are absent in pure numerical simulations. To bridge this gap for zero-shot transfer, we employ FDA to project the synthetic data into the frequency manifold of the target device.

The FDA process focuses on the spectral decomposition of the diagnostic signals. For a synthetic signal $x_{syn}$within a 1~ms window, we first apply the discrete Fourier transform (DFT) to obtain its frequency-domain representation:
\begin{equation}
\mathcal{F}[x_{\mathrm{syn}}] = A_{\mathrm{syn}}(f) \cdot e^{i\phi_{\mathrm{syn}}(f)}.
\label{eq:fda1}
\end{equation}
Here, $A_{\mathrm{syn}}(f)$ and $\phi_{\mathrm{syn}}(f)$ denote the amplitude and phase spectra, respectively. To minimize the domain shift between simulation and experiment, we extract the characteristic spectral signatures from the source device's experimental data. Specifically, we analyze 150~ms of disruptive precursor segments from EAST discharges (corresponding to 150 windowed samples), applying DFT to obtain their averaged amplitude spectrum, denoted as $\bar{A}_{\mathrm{src\_exp}}(f)$, which is then used to replace the amplitude spectrum of the J-TEXT synthetic diagnostic signals. The adapted signal is reconstructed by performing an inverse discrete Fourier transform while strictly preserving the original synthetic phase:
\begin{equation}
\hat{x}_{\mathrm{syn}\rightarrow \mathrm{tgt}} =
\mathcal{F}^{-1}
\left[
\bar{A}_{\mathrm{src\_exp}}(f)\cdot e^{i\phi_{\mathrm{syn}}(f)}
\right].
\label{eq:fda2}
\end{equation}
By performing this transformation, the resulting signal $\hat{x}_{\mathrm{syn}\rightarrow \mathrm{tgt}}$ retains the critical temporal precursors of MHD instabilities but exhibit the spectral distribution and noise characteristics of real-world discharge signals.

To evaluate the impact of FDA on feature alignment, we analyze the latent representations generated by the model's fusion layer. Using the previously fine-tuned TokaFormer, we perform forward inference on several datasets to extract the [CLS] tokens, i.e. the latent feature vectors utilized by the final classifier. The datasets include: (1) 5,000 samples from the final 25~ms of 200 disruptive J-TEXT pulses; (2) 25~ms of randomly sampled segments from each EAST test discharge; and (3) the original J-TEXT synthetic diagnostic signals.

\begin{figure}
    \centering
    \includegraphics[width=0.8\textwidth]{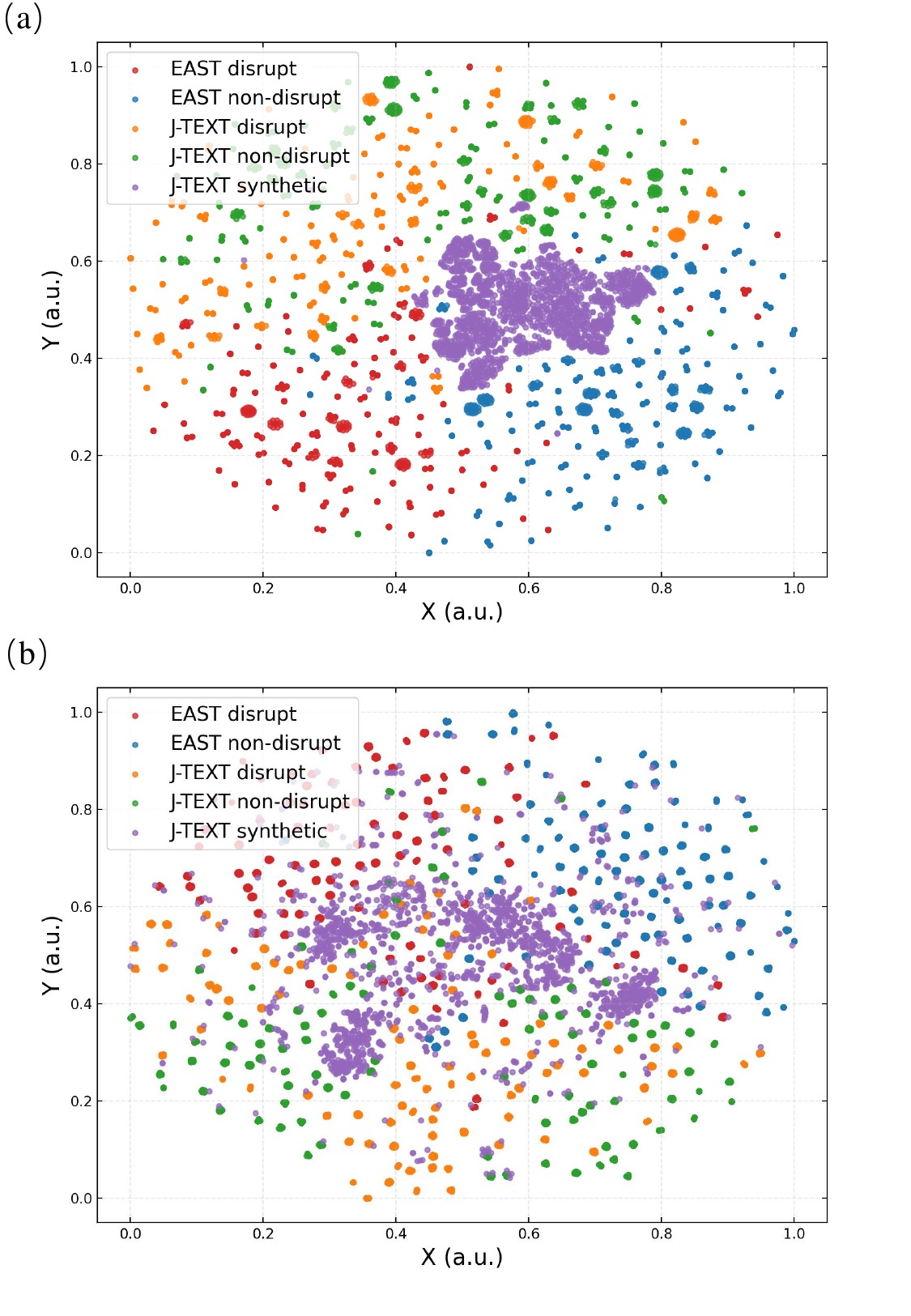}
    \caption{Comparison of latent feature t-SNE distributions before and after applying FDA. (a) The original synthetic signals (without FDA) form an isolated cluster, exhibiting a distinct domain shift from the target J-TEXT experimental data. (b) The FDA-adapted synthetic signals demonstrate successful feature alignment with the real-world J-TEXT distribution, proving the method's efficacy in bridging the sim-to-real gap.}
    \label{fig:tsne}
\end{figure}

We employ t-distributed stochastic neighbor embedding (t-SNE) to visualize the clustering of these feature vectors in a lower-dimensional space, as shown in Fig.\ref{fig:tsne}. The t-SNE analysis reveals that the original synthetic diagnostic signals occupy a distinct region of the feature space, failing to project onto the experimental J-TEXT data manifold. However, after applying FDA, the feature vectors of the adapted synthetic signals demonstrate a significant shift, mapping much more closely to the actual J-TEXT experimental distribution. This visualization confirms that spectral adaptation effectively aligns the simulation-based precursors with experimental realities, enabling the model to recognize disruptive patterns despite device-specific noise and diagnostic artifacts.

To further reduce the distribution discrepancy between the FDA-adapted synthetic diagnostic signals and the real experimental data from the target device, we incorporate CORAL during the transfer process. While FDA aligns the raw signal distributions in the spectral domain, CORAL minimizes the domain shift in the latent feature space by aligning the second-order statistics (covariance) of the source and target distributions.
In our implementation, we calculate the covariance distance between the features $D_s$ of the adapted synthetic domain and $D_t$ of the target experimental domain. The CORAL loss is defined as the squared Frobenius norm of the difference between the covariance matrices of the source and target features:
\begin{equation}
L_{\mathrm{CORAL}} =
\frac{1}{4d^2}
\left\|C_s - C_t\right\|_F^2,
\label{eq:coral}
\end{equation}
where $d$ is the feature dimension, and $C_s$ and $C_t$ represent the covariance matrices of the synthetic and experimental features, respectively. By minimizing this loss during the final alignment phase, we ensure that the latent manifolds of the two domains are statistically consistent.

Finally, we conduct zero-shot testing on 1,596 real experimental discharges from the J-TEXT device. The results, summarized in Table~\ref{tab:zeroshot}, demonstrate the significant performance gains achieved through this multi-stage alignment strategy. The results indicate that the combination of spectral adaptation and latent covariance alignment effectively bridges the simulation-to-reality gap. The final model achieves a 7\% improvement in overall accuracy compared to the initial transfer.
In a more detailed performance comparison, the inclusion of untreated synthetic diagnostic data causes the overall accuracy to drop to 0.39. This degradation primarily stems from the severe simulation-to-reality gap, which biases the model toward aggressively classifying stable phases as disruptive precursors. Consequently, although the TPR peaks at 0.92, it triggers an unacceptably high FPR of 0.73, making the model impractical for real-world operations.\\

More crucially, after applying the FDA spectral adaptation strategy, the cross-domain robustness of the model improves substantially. Among the 1,596 tested discharges comprising 296 disruptive and 1,300 non-disruptive shots, the FDA-adapted synthetic data introduces a reasonable trade-off. While the TPR over the 296 disruptive shots experiences a moderate decrease to 0.69, the performance on the 1,300 non-disruptive shots exhibits a major enhancement, with the FPR plunging from 0.73 down to 0.45. This remarkable 28\% relative reduction in false alarms demonstrates that the model is no longer confounded by idealized simulation features or device-specific experimental noise. Instead, the FDA alignment successfully shifts the model's focus toward more universal and robust physical precursors, significantly reducing the false alarm rate while maintaining high sensitivity to disruptive precursors.

\begin{table}[h]
\caption{Zero-shot performance on J-TEXT after FDA and CORAL alignment.}
\centering
\begin{tabular}{l c c c}
\hline
Experiment & TPR & FPR & Accuracy \\
\hline
Without synthetic signals & 0.79 & 0.56 & 0.50 \\
With synthetic signals & 0.92 & 0.73 & 0.39 \\
With FDA synthetic signals & 0.69 & 0.45 & 0.57 \\
\hline
\end{tabular}
\label{tab:zeroshot}
\end{table}

\section{Conclusions and discussion}
\label{summary}

In this study, we propose an experiment-free cross-device disruption prediction framework enabled by synthetic diagnostic data augmentation. To address the data scarcity in new fusion devices, we develop TokaFormer, a multi-modal dual-tower transformer that separately encodes Mirnov and SXR diagnostics to capture both microscopic MHD fluctuations and macroscopic plasma states. Self-supervised pretraining further improves the robustness and generalization capability of the model. The model is trained on EAST discharges and transferred to the J-TEXT tokamak without using any real target-device disruption data. To reduce the sim-to-real and cross-device gaps, FDA is applied to align spectral characteristics between synthetic and experimental signals, whereas CORAL alignment minimizes feature distribution shifts in the latent space.\\

Experimental results on 1,596 J-TEXT discharges demonstrate that the proposed framework achieves 57\% prediction accuracy and improves the early warning rate by 7\% compared with the baseline without synthetic diagnostics. These results indicate that the proposed framework provides a scalable solution for disruption prediction in future fusion devices such as ITER and DEMO, where real disruption data will be extremely limited during early operation stages.\\

Although this work has successfully demonstrated a feasible paradigm of employing synthetic diagnostics for experiment-free cross-device disruption prediction, the performance gains on the target device still leave room for further optimization. The primary bottleneck stems from the severe constraints on computational resources, since first-principles 3D non-ideal MHD simulations are exceptionally expensive and the scale of our current simulated dataset remains modest. Consequently, the scaling potential of this data-augmentation framework has not yet been fully exploited.\\

To overcome these bottlenecks in future work, we plan to substantially scale up both the volume and diversity of our simulation datasets. This may be achieved by, first, exploring a wider range of tokamak operational scenarios as initial discharge equilibrium profiles and, second, simulating an enriched spectrum of MHD instability events. Simultaneously, we aim to upgrade the ESDP framework by developing and integrating additional diagnostic modules that simulate universal cross-device experimental systems. Through these development, we expect to further unlock the potential of this simulation-assisted approach and further enhance the predictive robustness for next-generation fusion reactors.

\section*{Acknowledgements}

We are grateful for the supports from the NIMROD team and the helpful suggestions from Prof. Zhoujun Yang and Prof. Zhipeng Chen. This work is supported by the National MCF Energy R\&D Program of China under Grant No.~2019YFE03050004 and the National Natural Science Foundation of China Grant No.~51821005. The computing work in this paper is supported by the Public Service Platform of High Performance Computing by Network and Computing Center of HUST.

\section*{References}
\bibliography{TokaFormer}

\clearpage
% \begin{figure}
%     \centering
%     \includesvg[width=0.8\textwidth]{images/pipeline.svg}
%     \caption{Overview of the proposed zero-shot cross-device transfer framework for tokamak disruption prediction. The framework contains three stages: (1) source-domain training using the TokaFormer dual-tower architecture with Mirnov and SXR diagnostics; (2) synthetic target-domain signal generation through NIMROD MHD simulation and the enhanced synthetic diagnostic platform (ESDP) with domain randomization; and (3) zero-shot transfer to the target device using FDA-based spectral adaptation and CORAL-based feature alignment. The pretrained backbone is frozen during transfer, enabling direct inference on real target-domain discharges without target-device training data.}
%     \label{fig:Pipeline}
% \end{figure}

\clearpage
\begin{figure}
    \centering
    \includegraphics[width=\textwidth]{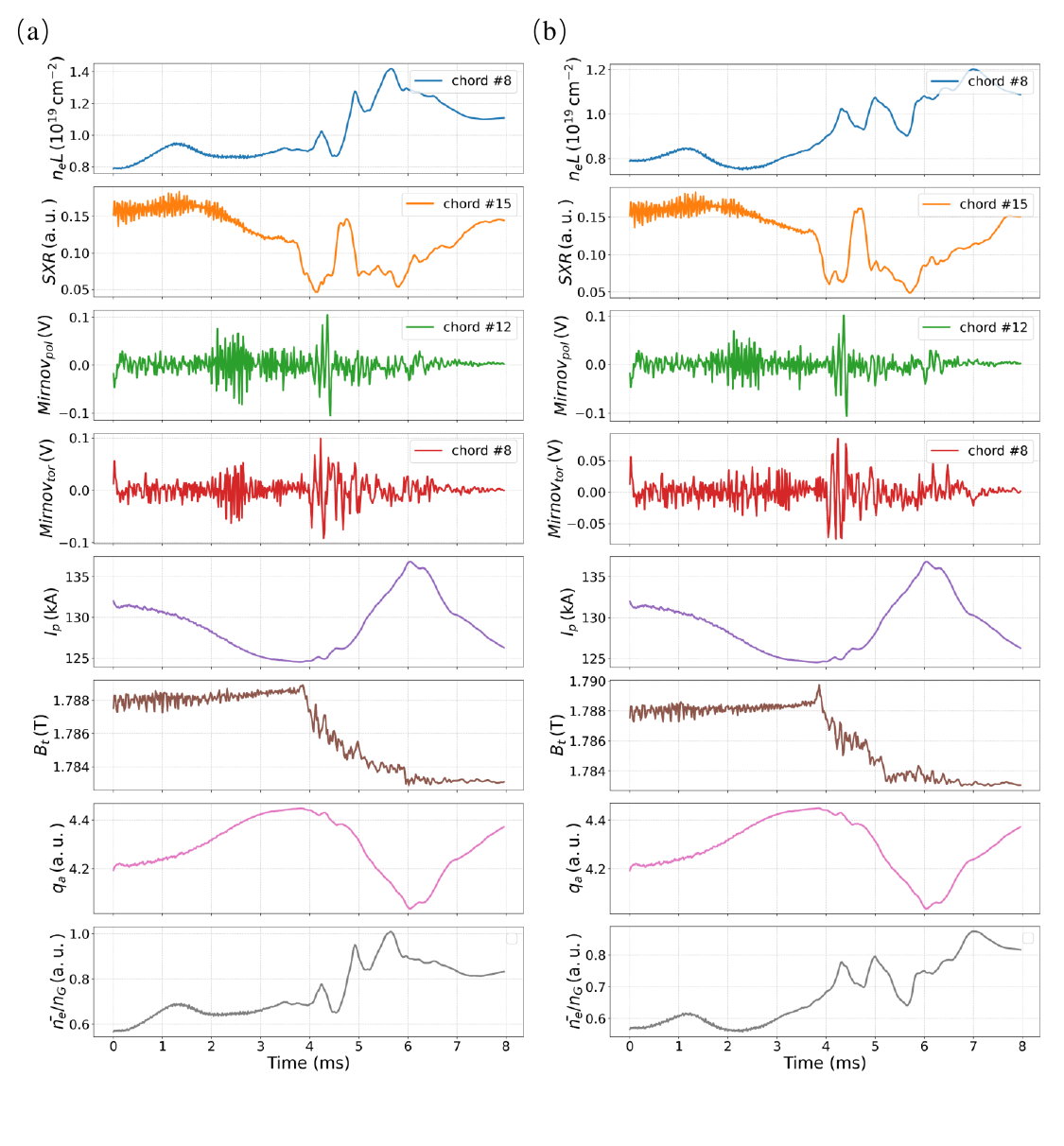}
    \caption{Synthetic diagnostic signals as functions of time at toroidal angles: (a) $0^\circ$ and (b) $30^\circ$ as derived from 3D MHD simulations for the line integrated of electron number density $n_eL$, soft X-ray signal, Mirnov signals from poloidal and toroidal arrays, plasma current $I_p$, toroidal magnetic field $B_t$, safety factor $q_q$ at plasma boundary, and the center line-averaged electron number density normalized by the Greenwald density limit (from top to bottom panels respectively).}
    \label{fig:synthetic_toroidal}
\end{figure}

\clearpage

\clearpage

\clearpage

\clearpage

\end{document}